\documentclass[a4paper,11pt]{article}
\usepackage[latin2,utf8]{inputenc}
  \usepackage{xcolor}

\usepackage{amsmath,amssymb,amsthm, amsfonts}
\usepackage{verbatim}
\usepackage{graphicx}
\usepackage{arydshln}

\numberwithin{equation}{section}
\numberwithin{figure}{section}
\numberwithin{table}{section}

\begin{document}

\begin{center}
{\bf A new SEIR type model including quarantine effects
and its application to analysis of Covid-19 pandemia in Poland in March-April 2020}

\bigskip

{\sc Tomasz Piasecki$^1$, Piotr B. Mucha$^1$,
Magdalena Rosi\'nska$^2$}

\end{center}

\noindent
1. Institute of Applied Mathematics and Mechanics, University of Warsaw,\\
ul.  Banacha 2, 02-097 Warszawa, Poland.\\
E-mails: t.piasecki@mimuw.edu.pl, p.mucha@mimuw.edu.pl

\smallskip

\noindent
2. Department of infectious Disease epidemiology and Surveillance,\\
National Institute of Public Health - National Institute of Hygiene,\\
ul. Chocimska 24, Warsaw, Poland.  
\\
E-mail: mrosinska@pzh.gov.pl\\

\smallskip


\begin{abstract}

    Contact tracing and quarantine are well established non-pharmaceutical epidemic control tools. The paper aims to clarify the impact of these measures in COVID-19 epidemic. 
    A new deterministic model is introduced (SEIRQ: susceptible, exposed, infectious, removed, quarantined) with Q compartment capturing individuals and releasing them with delay. We obtain a simple rule defining the reproduction number $\mathcal{R}$ in terms of 
    quarantine parameters, ratio of diagnosed cases and transmission parameters. The model is applied to the epidemic in Poland in March - April 2020, when social distancing measures were in place. We investigate 3 scenarios corresponding to different ratios of diagnosed cases. Our results show that depending on the scenario contact tracing could have prevented from 50\% to over 90\% of cases. The effects of quarantine are limited by fraction of undiagnosed cases. Taking into account the transmission intensity in Poland prior to introduction of social restrictions it is unlikely that the control of the epidemic could be achieved without any social distancing measures.

\end{abstract}

\bigskip


\smallskip

\section{Introduction}
The epidemic of SARS-CoV-2 infection triggered an unprecedented public health response. 
Given the lack of effective vaccine and treatment in 2020, this 
response included a variety of travel restrictions and social distancing measures \cite{Tian}. 
While these measures help to slow down the epidemic they come at significant economical and societal 
cost \cite{ECDCRRA}.
As an alternative an approach focusing on rapid diagnosis is increasingly recommended \cite{WHO.S} and prior to lifting social distancing measures
large-scale community testing should be in place \cite{EC}. Testing efforts are complemented by identifying and quarantining contacts of the diagnosed cases. Of note, by isolating the asymptomatic contacts from their social networks, this strategy takes into account the pre-symptomatic and asymptomatic spread of the infection \cite{Tong, Huang}, believed to be one of the key drivers of fast spread of COVID-19.
As an example, wide spread testing in general population followed by isolation of the infected 
helped to reduce COVID-19 incidence by 90\% in an Italian village of Vo’Euganeo \cite{Day}. 
A modelling study in France offers similar conclusions arguing that relaxing social lock-down will 
be only feasible in case of extensive testing \cite{Di}.
While there is already a number of studies estimating the effects of general social distancing measures \cite{Flaxman, Tian, Di, Giordano}, less is known about the impact of quarantine. Hellewell et al. \cite{Hellewell} investigated the potential of rapid isolation of cases and contact tracing to control the epidemic, finding that prohibitively high levels of timely contact tracing are necessary to achieve control. However, new technologies may offer sufficiently fast alternative to traditional contact tracing, in which case the epidemic could be still controlled by contact tracing \cite{Ferretti}.   

\smallskip

Our aim is to develop a SEIR-type model which incorporates the effects of quarantine and validate it in a setting in which measures to reduce contacts are in place. We apply it to investigate the role of quarantine in Poland.
The first case of COVID-19 in Poland was diagnosed on March 4th. Social distancing measures were rapidly introduced during the week of 9 - 13th March including closure of schools and universities, cancellation of mass events and closure of recreation facilities such as bars, restaurants, gyms etc. as well as shopping malls. Religious gatherings were limited. Finally, borders were closed for non-citizens \cite{Pinkas}. These measures were fully in place on March 16th. Further, beginning at March 25th restrictions on movement and travel were introduced (lock-down). Wearing face covers became obligatory on April 14th. The restrictions were gradually lifted beginning at April 20th. 
We focus on modelling the time period when the social distancing measures were in place and then consider different scenarios of relaxation of the restrictions with possible improvement of testing and contact tracing. We note that the procedures for quarantine were in place even before the social distancing measures. They initially focused on individuals arriving from COVID-19 affected areas in China. When the epidemic started spreading in European countries people who came back to Poland from these countries were advised to immediately seek medical attention if they experienced any symptoms consistent with COVID-19. However, adherence to these recommendations was not evaluated. As soon as the first case was diagnosed in Poland quarantine for close contacts was also implemented.

\smallskip

This paper aims to define a deterministic population model describing the epidemic in classical terms of susceptible, exposed, infectious, removed.
In our model the quarantine becomes a separate state that removes individuals from susceptible and exposed states. We show that the reproductive number in our model is given by a simple formula referring to the parameters of transmission and transition, but also to parameters describing the quarantine.   
We demonstrate that in a real life scenario (case study of Poland) the quarantine effectively reduces the growth of infectious compartment. Increasing the efficiency of contact tracing and testing may may to some extent compensate lifting up the social distancing restrictions. 

\section{Methods}

\subsection{The model}
We introduce a modification of the classical SEIR model 
including effects of quarantine. To underline importance of that extension we call it SEIRQ. Formally the model is described by a system of ordinary differential equations with delay dedicated to the quarantine.

\begin{figure}[h!] \label{model} 
\centering 
\includegraphics[width=0.75\textwidth]{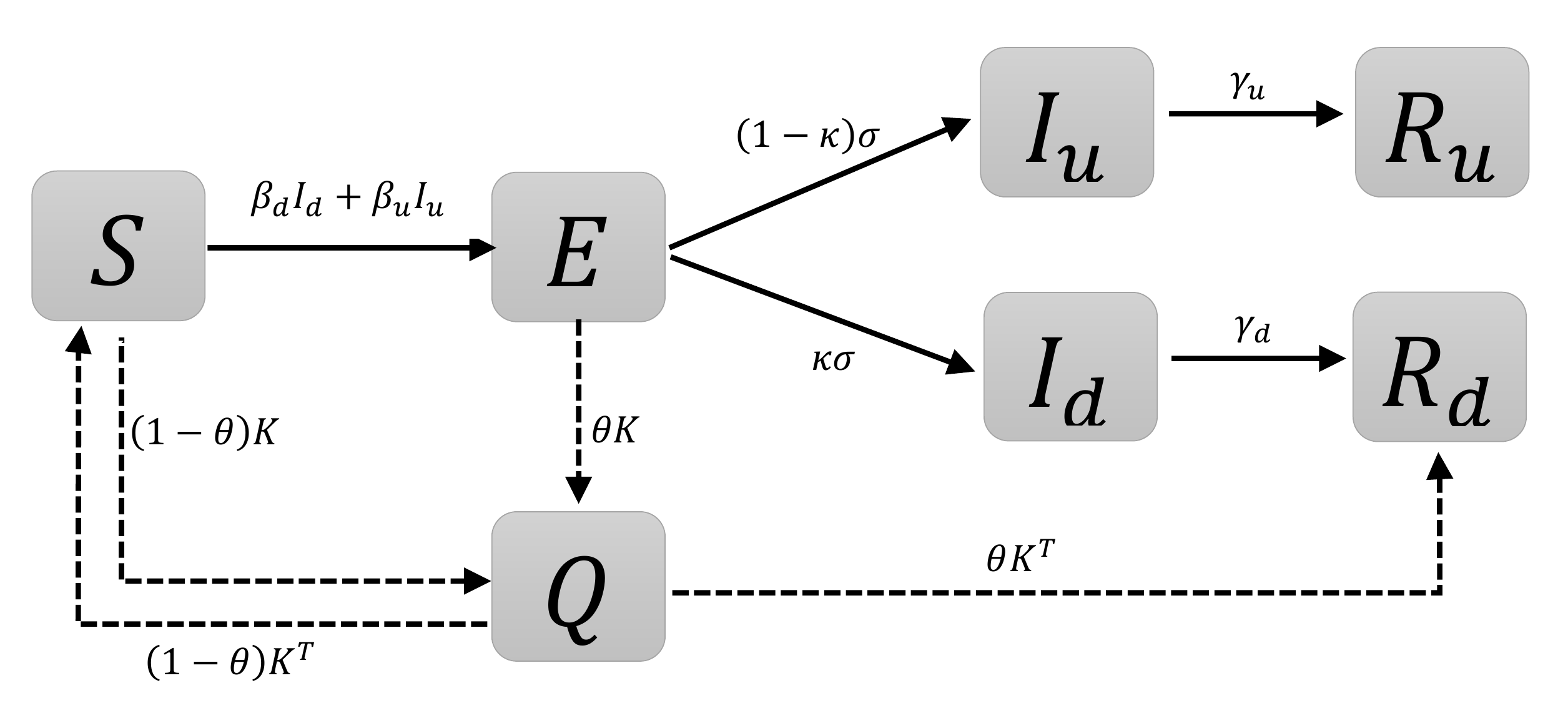}
\caption{{\bf Schematic representation of the states included in the model}. The solid lines represent the transition parameters and the dashed line indicate that the specific quantity is added.}
\end{figure} 

\noindent
The following states are included in the model:

$S(t)$ -- susceptible

$E(t)$ -- exposed (infected, not infectious)

$I_d(t)$ -- infectious who will be diagnosed 

$I_u(t)$ -- infectious who will not be diagnosed

$R_d(t)$ -- diagnosed and isolated

$R_u(t)$ -- spontaneously recovered without being diagnosed

$Q(t)$ -- quarantined

\bigskip

The figure \ref{model} 
presents the schematic representation of the model. A susceptible individual (state $S$), when becoming infected first moves to the state $E$, to model the initial period, when the infected individual is not yet infectious. Next the cases progress to one of the infectious states $I_d, I_u$ at the rates $\kappa\sigma$ and $(1-\kappa)\sigma$, respectively. In general, moving through the $I_d$ pathway concerns these individuals who (independently of quarantine) would meet the testing criteria, as relevant to the local testing policy, e.g. testing of people with noticeable symptoms.
We shall emphasize that from the point of view of analysis of spread of infection, the quantity $I_u$
shall be regarded rather as not recognized infections, not necessarily asymptomatic or mild. 
With this interpretation the value of $\kappa$ can be influenced by intensity of testing. 

The creation of state $E$ is via $I_d$ and $I_u$ with transmission rates $\beta_d$ and $\beta_u$, respectively, normalized to the total population size $N=S+E+I_d+
I_u+R_d+R_u+Q$, which is assumed to be constant in time, births and deaths are neglected.

The transition parameter $\sigma$ is assumed identical for both groups, relating to the time between infection and becoming infectious. The infectious individuals then move to the state $R_d$, which is the state of being diagnosed and isolated (and later recovered or deceased), with the rate $\gamma_d$ corresponding to the observed time between onset and diagnosis. On the other hand $R_u$ contains people who spontaneously recovered with rate $\gamma_u$. 
Our model includes an additional state of being quarantined ($Q$). To mimic the situation of contact tracing, individuals can be put in quarantine from the state $S$ (uninfected contacts) or the state $E$ (infected contacts). These individuals stay in the quarantine for a predefined time period $T$.  We assume that the number of people who will be quarantined depend on the number of individuals who are diagnosed.
An average number of individuals quarantined per each diagnosed person is denoted as $\alpha$. However, as the epidemic progresses some of the contacts could be identified among people who were already infected, but were not previously diagnosed, i.e. the state $R_u$. We note that moving individuals between the states $Q$ and $R_u$ has no effect on the epidemic dynamics, therefore we assume that only individuals from $S$ and $E$ are quarantined and we reduce the average number of people put on quarantine by the factor $\frac{S(t)}{S(t)+R_u(t)}$. Further, to acknowledge the capacity limits of the public health system to perform the contact tracing, we introduce a quantity $K_{max}$, describing the maximum number of people who can be put in quarantine during one time step.  

We also assume that among the quarantined a proportion $\theta$ is infected. After the quarantine, the infected part $\theta K(t-T)$ goes to $R_d$ and the rest $(1-\theta)K(t-T)$ returns to $S$. 

Taking all of the above into account, the model is described with the following SEIRQ system:
\begin{equation}\label{seir}
 \begin{array}{l}
  \dot S(t)=- S(t) (\beta_d I_d(t)+\beta_uI_u(t)) - (1-\theta)K(t) + (1-\theta)K(t-T), \\[5pt]
  \dot E(t)=  S(t) (\beta_d I_d(t) + \beta_u I_u(t)) - \sigma E(t)  - \theta  K(t)\\ [5pt]
  \dot I_d(t)=\kappa \sigma E(t) -   \gamma_d I_d(t), \\[5pt]
  \dot I_u(t)=(1-\kappa) \sigma E(t) -\gamma_u I_u(t),\\[5pt]
  \dot R_d(t)= \gamma_d I_d(t) + \theta K(t-T),\\ [5pt]
  \dot R_u(t)=\gamma_u I_u(t),\\[5pt]
  \dot Q(t)= K(t)-K(t-T),\\[5pt]
  \mbox{where \ \ } 
  K(t)={\rm min}\{\frac{S(t)}{S(t)+R_u(t)} \alpha \gamma_d I_d(t),  K_{max}  \},
  \mbox{\ \ and  \ \ } \alpha,\beta_d,\beta_u,\gamma_d,\gamma_u, \theta, T \geq 0.
 \end{array}
\end{equation}
We assume that the parameters $\alpha, \beta_d, \beta_u, \theta$, $\gamma_u$ and $\gamma_d$  depend on the country and time-specific public health interventions and may therefore change in time periods. Due to proper interpretation of the equation on $E$ we require that $\beta_d \geq 
\theta \alpha \gamma_d$ to ensure positiveness of $E$.

\subsection{Basic reproductive number, critical transmission parameter $\beta^*$.} 

Based on the general theory of SEIR type models \cite{Dik}, we introduce the reproductive number
\begin{equation} \label{def:R0}
 \mathcal{R}=\kappa \left( \frac{ \beta_d}{\gamma_d} - \theta \alpha \right)  + (1-\kappa) \frac{\beta_u}{\gamma_u}.
\end{equation}
It determines the stability of the system as $\mathcal{R} <1$
and instability for $\mathcal{R}>1$ (the growth/decrease of pandemia).
This quantity not only explains the importance is testing (in terms of $\kappa$) and
quarantine (in terms of $\alpha$), but also gives an indication on levels of optimal testing and contact tracing. We underline that this formula works for the case when the capacity of the contact tracing has not been exceeded $(K(t)<K_{max})$.
The details of derivation of \eqref{def:R0} are provided in the Appendix, section \ref{sec:R}. We shall emphasize the formal mathematical derivation holds for the case when $I$ and $E$ are small comparing to  $S$. Therefore the complete dynamics of the nonlinear system (\ref{seir}) is not fully determined by \eqref{def:R0}. However in the regime of epidemic suppression, which is the case of COVID-19 epidemic in Poland, $I$ and $E$
are small compared to $S$ and so the formula (\ref{def:R0})
reasonably prescribes spreading of 
infection in the population.

The critical value $\mathcal{R}=1$ defines the level of transmission which is admissible, taking into account the existing quarantine policy, in order to control epidemic. As the level of transmission depends on the level of contacts, this provides information on the necessary level of social distancing measures.
The formula \eqref{def:R0} indicates that improving the contact tracing may compensate relaxation of contact restrictions. The key quantity is $\theta \alpha$. Indeed the system with the quarantine 
has the same stability properties as one without  $K$, but with the  new 
transmission rate $\beta_d^{new}=\beta_d - \theta \alpha \gamma_d$. 
In order to guarantee the positiveness of $E$, $\beta_d^{new}$ must be nonnegative.
It generates the constraint 
\begin{equation} \label{3.1}
\theta\alpha\gamma_d \leq \beta_d.
\end{equation}
%
The above condition also implies the theoretical maximal admissible level of quarantine. We define it by improving the targeting of the quarantine, i.e. by the highest possible level of $\theta$:
\begin{equation}\label{def:thetamax}
\theta_{max} = \frac{\beta_d}{\gamma_d\alpha}.
\end{equation} 
As long as the $K_{max}$ threshold is not exceeded the effect of the increase in $\theta$ or in $\alpha$ play the same role at the level of linearization (small $I, E$). However, in general it is not the case and for the purpose of our analysis we fix $\alpha$.

For our analysis we assume $\beta_d=\beta_u=\beta$.
The reason is that, both $I_d$ and $I_u$  contain a mixture of asymptomatic and symptomatic cases and although there might be a difference we lack information to quantify this difference.
%
%
Then
using formula \eqref{def:R0} we  compute critical values $\beta^*(\kappa,\theta,\alpha)$ defined as 
\begin{equation} \label{def:betacrit}
{\mathcal R}(\beta^*)=1, \mbox{  \ namely \ }
\beta^*(\kappa,\theta,\alpha,\gamma_d,\gamma_u)=\frac{(1+\theta\alpha\kappa)\gamma_d\gamma_u}{\gamma_u\kappa+\gamma_d(1-\kappa)}.
\end{equation}
It shows the upper bound on transmission rate $\beta$ which still guarantees the suppression of pandemic. We shall omit the dependence on $\gamma_d,\gamma_u$ as these are fixed in our case, and denote briefly $\beta^*(\kappa,\theta,\alpha)$.

In the case of maximal admissible quarantine (\ref{def:thetamax})
 we obtain
\begin{equation} \label{betamax}
 \beta^*(\theta_{max},\kappa) = \frac{\gamma_u}{1-\kappa},
\end{equation}
which can be regarded as theoretical upper bound for $\beta$ if we assume "optimal admissible" quarantine
for fixed $\kappa$, for which the epidemic could be still controlled. 
It must be kept in mind though that the condition (\ref{3.1}) means that we are able to efficiently isolate all persons infected by every diagnosed, therefore is unrealistic. The resulting $\beta^*(\theta_{max},\kappa)$ should be therefore
considered as a theoretical limit for transmission rate.

\subsection{Fitting procedure} 
All simulations are performed using GNU Octave ({\em https://www.gnu.org/software/octave/}). 
The underlying tool for all computations is a direct finite difference solver with a 1 day time step. 

\smallskip

{\bf Basic assumptions for data fitting.}
We estimate the transmission rates $\beta$ by fitting the model predictions to the data on the cumulative number of confirmed cases.
%
%
%
%
Since people with confirmed diagnosis are efficiently isolated,
  they are immediately included into $R_d$. Therefore,
the quantity fitted to the data is $R_d(t)$. 

The crucial assumption behind our approach is that the parameter $\beta$ changes twice 
during the period of analysis. The reason is that we can distinguish two important time points in the development of 
epidemic in Poland. The first is initial restrictions including school closure effective March 12, which was accompanied with restrictions on other social activities.
As we do not take migration into account in our model, we  assume that the effect of border closing is reflected in $\beta$. The second turning point was a lockdown announced on March 25.    
  
For simplicity we comprise the effect of above measures in two jump changes in $\beta$ in $t \in \{t_1,t_2\}$
and choose $t_1=14, \quad t_2=28$. 
With $t=1$ corresponding to March 3 it means small delay with respect to the above dates which can be justified 
by the fact that new cases are reported with a delay of approximately 2 days.


\smallskip

{\bf Choice of fixed parameters (Tab. \ref{tab0}).}
We assume that the parameters $\sigma, \gamma_u$ represent the natural course of infection and their values could be based on the existing literature. The parameter $\sigma$ describes the rate of transition from non-infectious incubation state $E$ into the infectious states $I_d$ or $I_u$. The value of $\sigma$ takes into account the incubation period and presymptomatic infectivity period.
 $\gamma_u$ relates to the period of infectivity, which we select based on the research regarding milder cases, assuming that serious cases are likely diagnosed. Further, $\kappa$ is a parameter related both to the proportion of asymptomatic infection and the local testing strategies. Since the literature findings provide different possible figures, for $\kappa$ we examine three different scenarios.

Parameters $\gamma_d, \theta$ and $\alpha$ are fixed in our model for the purpose of data fitting, but informed by available data. One of the scenarios of future dynamics of the epidemic (section \ref{sec:future}) considers possible increase of $\theta$. Parameter $\gamma_d$ was estimated basing on time from onset to diagnosis for diagnosed cases, and $\theta$ as rate of diagnosed among quarantined. Furthermore we  fix the parameter $\alpha$ by comparing the number 
of quarantined people obtained in simulations with actual data. 
The capacity level of public health services is set in terms of possible number of quarantined per day $K_{max}$, as double the level observed so far. Detailed justification of the values of fixed parameters collected in the following table, is given in the Appendix, section \ref{sec:par}. 
\begin{table}[h!] 
\centering
\begin{tabular}{|c|c|c|}  
\hline
& &  \\[-8pt]
Parameter & Value & Source \\[4pt]
\hline& &  \\[-8pt]
$\sigma$    & $\frac{1}{3.5}$ & Literature: incubation time \cite{Guan, Li.Q, Lauer} \\
& & 
+ presymptomatic spread \cite{Wei, Tong, Huang}   \\[4pt]
\hline
& &  \\[-8pt]
$\gamma_d$  & $\frac{1}{5.5}$ &  Observed data: appendix section \ref{sec:par}  \\[4pt]
\hline
& &  \\[-8pt]
$\gamma_u$  & $\frac{1}{10}$  &  Literature: \cite{Hu}, WHO mission report from China  \\[4pt]
\hline
& &  \\[-8pt]
$\kappa$    & \{0.2; 0.5; 0.8\} & Literature: proportion asymptomatic\\
& &
or undocumented \cite{Li.R, Nishiura, Day,John} \\[4pt]
\hline
& &  \\[-8pt]
$\theta$    &  0.006 & Observed data: appendix section \ref{sec:par} \\[4pt]
\hline
& &  \\[-8pt]
$\alpha$  & 75 & Observed data: appendix section \ref{sec:par} \\[4pt]
\hline
& &  \\[-8pt]
$K_{max}$  & 50 000 & $2 \,\times$ the maximum level observed so far (arbitrary decision) \\[4pt]
\hline
\end{tabular} 
\caption{Fixed parameters used in the model}
 \label{tab0}
\end{table}

{\bf Optimization algorithm.} 
In order to fit the values $\beta_1,\beta_2,\beta_3$ we use a standard gradient descent algorithm. The error function is defined as mean square difference between the cumulative number of diagnoses and the $R_d(t)$ predicted from the model.

For the initial values the error function is optimized only for a limited number of possible conditions, as these mostly impact $\beta_1$, which is less relevant for future predictions.
To estimate confidence intervals we use a method of parametric bootstrap. The optimisation procedures are described in the Appendix, section \ref{sec:opt}, where we also show precise errors of data fitting.

\smallskip
{\bf Dataset.} 
 The data series contains cumulative number of confirmed cases of COVID-19 in Poland  
from March 3 (first confirmed case in Poland) till April 26, which amounts 
to 54 observations. The data are taken from official communications of the Ministry of Health.
As explained in table \ref{tab0} and appendix (section \ref{sec:par} additional data sources were used for choosing $\theta$, $\alpha$ and $\gamma_d$.

\section{Results} \label{s:sym}
\subsection{Estimation of parameters  and "no-change" scenario predictions}
\begin{figure}[h!]
	\begin{center}
	\includegraphics[width=12cm]{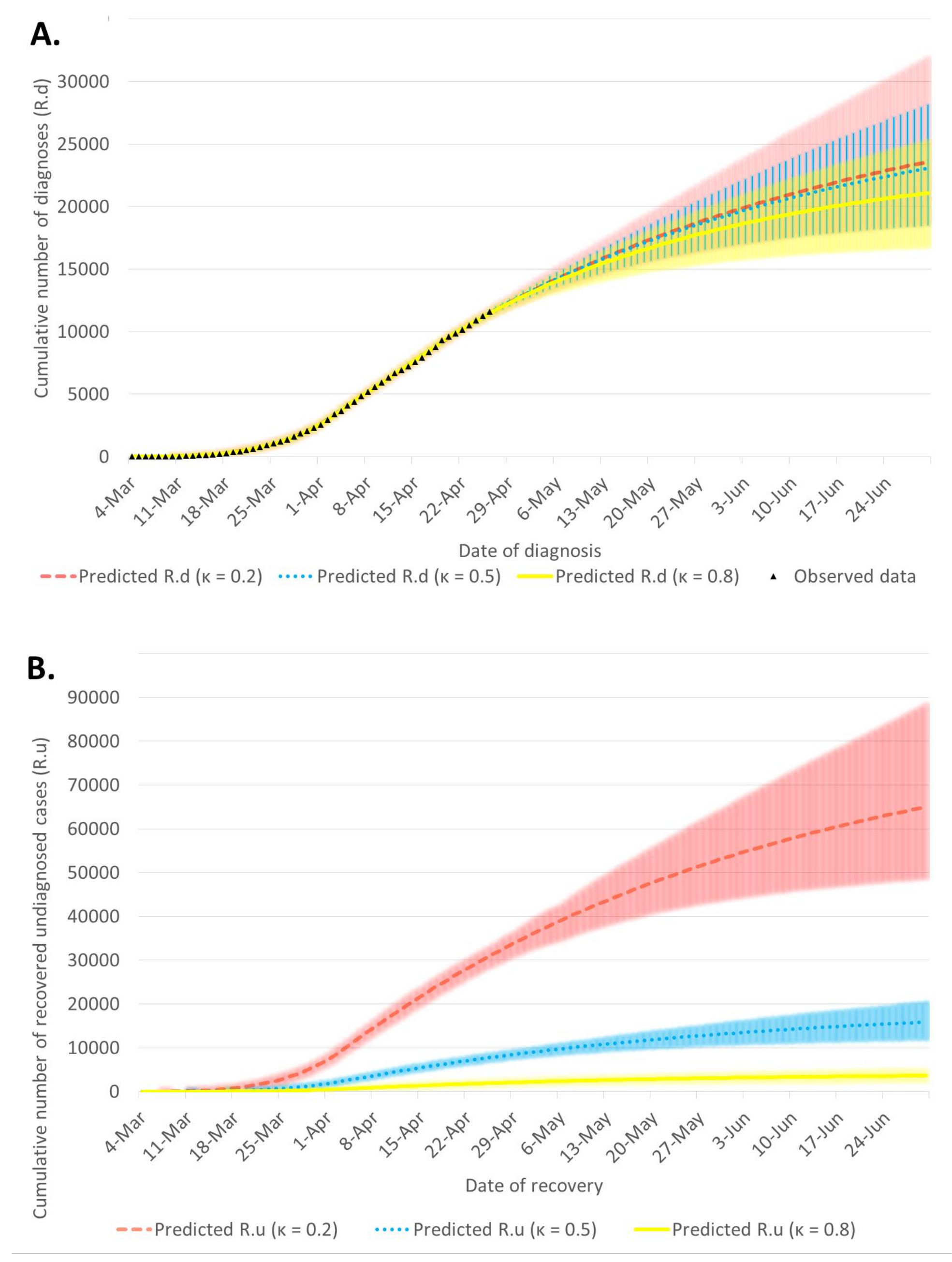}
	\end{center}
\caption{Results of model fit to cumulative diagnosed cases ($R_d$) for $\kappa=0.2,0.5,0.8$ (panel A) and corresponding predictions for undiagnosed, recovered compartment, $R_u$ (panel B). Coloured shades correspond to 95\% confidence intervals for the respective colour line.}	
\label{fig2.0}
\end{figure}

In Table \ref{tab2.1} we show estimated values of $\beta_i$, where $i=1,2,3$ correspond to the time intervals when different measures were in place, and the ${\mathcal R}$ for the third time interval. Given the social distancing measures in place early April 2020, as well as the quarantine levels, the reproductive number was below 1, independently of the value of $\kappa$, which relates to testing effectiveness. 
The figure \ref{fig2.0} shows the fit of the models assuming different levels of $\kappa$. Good fit is found for all three models although predictions start to differ in the middle-term prognosis.

\begin{table}[h!]
\centering
\begin{tabular}{|c|c|c|c|}
\hline
& & & \\[-8pt]
& $\kappa=0.2$ & $\kappa=0.5$  & $\kappa=0.8$ \\[4pt]
\hline
& & & \\[-8pt]
$\beta_1$ & 0.635 & 0.684 & 0.738 \\[4pt]
        & (0.569 , 0.701) & (0.611 , 0.744) & (0.672 , 0.812) \\[4pt]
\hdashline
& & & \\[-8pt]
$\beta_2$ & 0.332 & 0.383 & 0.442 \\[4pt]
        & (0.288 , 0.397) & (0.336 , 0.443) & (0.4 , 0.514) \\[4pt]
\hdashline
& & & \\[-8pt]
$\beta_3$ & 0.099 & 0.132 & 0.175 \\[4pt]
        & (0.081 , 0.118) & (0.11 , 0.149) & (0.147 , 0.214) \\[4pt]
\hdashline
& & & \\[-8pt]        
$\mathcal{R}(\beta_3, 0.006,75)$ & 0.817 & 0.802 & 0.772 \\[4pt]
        & (0.651 , 0.977) & (0.648 , 0.915) & (0.569 , 0.874) \\[4pt]
\hline
\end{tabular} 
\caption{Estimated values of $\beta_i$ and values of $\mathcal{R}$ corresponding to the latest estimation period with 95\% confidence intervals }
 \label{tab2.1}
\end{table}

\begin{figure}[h!]
	\centering
	\includegraphics[width=16cm]{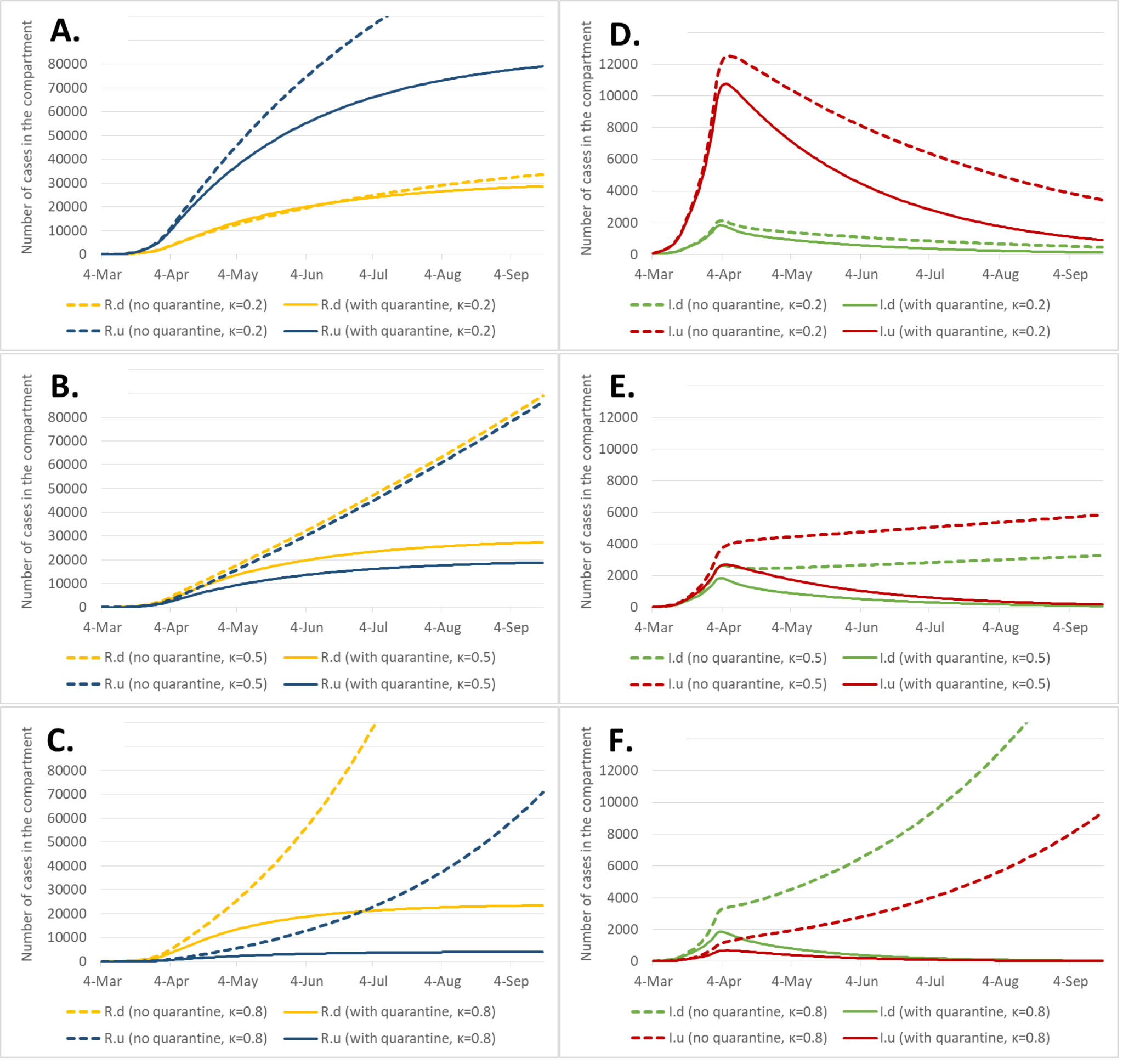}
\caption{Predicted values of $R_d,R_u$ (panels A -- C)  and $I_d, I_u$ (panels D -- E), as depending on the value of $\kappa$ and whether or not the quarantine is implemented. For $t>54$ $\beta=\beta_3$ estimated for each $\kappa$, with the same quarantine parameters or without quarantine at all.}
\label{fig2.2a}
\end{figure}

We proceed with predictions assuming that the restrictions are continued, i.e.keeping $\beta=\beta_3$ (note that the estimated $\beta_3$ is different for each $\kappa$). We calculate the epidemic duration ($t_{max}$), the peak number of infected ($I_d^{max}, I_u^{max}$) and the final size of the epidemic ($R_d(t_{max}), R_u(t_{max})$).  
In order to show the influence of quarantine we compare the situation with quarantine, keeping the same $\theta,\alpha$, to the situation without quarantine, setting $\alpha\theta=0$. The results of the development of the epidemic during the first 120 days are shown on Figure \ref{fig2.2a}.


For $\kappa=0.2$ the difference between the scenarios with and without quarantine is visible but not striking. However for  $\kappa=0.5$ and $\kappa=0.8$ a bifurcation in the number of new cases occurs around $t=40$ leading to huge difference in the total time of epidemic and total number of cases. These values are summarized in the table \ref{tab2.2}. We note that given the epidemic state in the first half of April 2020 for all values of $\kappa$ the model predicts epidemic extinction both with quarantine and without quarantine.
However, since the epidemic is very near to the endemic state, the predicted duration is very long, especially if no quarantine is applied.

\begin{table}[h!]
\centering
\begin{tabular}{|c|c|c|c|c|c|c|}
\hline
& & & & & & \\[-8pt]
$\kappa$ & quarantine factors & $R_d(t_{max})$ & $R_u(t_{max})$ & $I_d^{max}$ & $I_u^{max}$ & $t_{max}$ \\ 
[4pt]
\hline 
& & & & & & \\[-8pt]
0.2 & $\theta=0.006, \alpha=75$ & 31 & 85 & 1.9 & 10.8 &450 \\
& $\theta,\alpha = 0$ & 44 & 175 & 2.1 & 12.5 & 830 \\[4pt]
\hline
& & & & & & \\[-8pt]
0.5 & $\theta=0.006, \alpha=75$ & 29 & 20 & 1.9 & 2.7 & 330 \\
& $\theta,\alpha = 0$ & 1078 & 1078 & 5.1 & 9.2 & 3200 \\[4pt]
\hline
& & & & & &  \\[-8pt]
0.8 & $\theta=0.006, \alpha=75$ & 24 & 4 & 1.9 & 0.7 & 230 \\
& $\theta,\alpha = 0$ &  6317  & 1579 & 10.6 & 47.6 & 1280 \\[4pt]
\hline
\end{tabular} 
\caption{ Duration of epidemic ($t_{max}$) in days, the final values of $R_d$ and $R_u$, in thousands ($R_d(t_{max}), R_u(t_{max})$) and peak values of $I_d$ and $I_u$, in thousands ($I_d^{max}, I_u^{max}$) according to quarantine and testing scenarios.}
\label{tab2.2}
\end{table}

%




\subsection{Critical $\beta^*$ for the current situation}
Using the formula \eqref{def:betacrit} we can compute critical values $\beta^*$. 
In Table \ref{tab1} we show the values of $\beta^*(\kappa,0.006,75)$ and for convenience recall 
also estimated values of $\beta_3$ and ${\mathcal R}$, listed already in Table \ref{tab2.1}. 
Moreover we compute $\beta^*(\kappa,0,0)$, i.e. without quarantine and show values of ${\mathcal R}$
for our estimated values of $\beta_3$ and the same $\gamma_d, \gamma_u$ but without quarantine. Comparing the estimated values of $\beta_3$ (table \ref{tab1}) 
for all cases of $\kappa$ are only slightly below $\beta^*$.  
\begin{table}[h!]
\centering
\begin{tabular}{|c|c|c|c|c|c|c|}  
\hline
$\kappa$ & $\beta_3$ & $\beta^*(\kappa,0.006,75)$  & ${\mathcal R}(\beta_3,\kappa,0.006,75)$ & 
$\beta^*(\kappa,0,0)$ & ${\mathcal R}(\beta_3,\kappa,0,0)$ & $\beta_3 - \theta \alpha \gamma_d$ \\
\hline
0.2 & 0.099 & 0.12 & 0.817 & 0.11 & 0.907 & 0.018 \\ 
0.5 & 0.132 & 0.158 & 0.802 & 0.129 & 1.03 & 0.051 \\
0.8 & 0.175 & 0.211 & 0.772  & 0.155 & 1.132 & 0.074 \\
\hline
\end{tabular} 
\caption{Values of $\beta^*$ and $\mathcal R(\beta_3)$ with 
quarantine ($i=0.006, \; \alpha=75$) and without quarantine }
\label{tab1}
\end{table}

Eliminating the quarantine, for the estimated values of $\beta_3$, we have different 
situations depending on the actual value of $\kappa$. In case $\kappa=0.2$, so assuming that currently only 20\% of infections are diagnosed, the low values of ${\mathcal R}$ are due to low $\beta_3$ rather than the effect of quarantine (controlling epidemic by social contact restrictions). In effect even if we remove the quarantine we have still ${\mathcal R}<1$, but very close to 1. On the other hand if $\kappa=0.5$ or $\kappa=0.8$ we estimate higher $\beta_3$, which corresponds to the situation of controlling the epidemic by extensive testing and quarantine. For these cases, if we remove the quarantine, we end up with ${\mathcal R}>1$. 
The quantity $\beta_3-\theta\alpha\gamma_d$ represents effective transmission rate due to diagnosed cases. In particular it shows by how much the transmission could be reduced by improved contact tracing ($\theta\alpha$) and faster diagnosis ($\gamma_d$).

These results confirm
that the higher is the ratio of undiagnosed infections, 
the weaker is influence of quarantine. In the next section we verify these results numerically.

\subsection{Impact of quarantine at relaxation of social distancing} \label{sec:future}
\begin{figure}[h!]
	\centering
	\includegraphics[width=14cm]{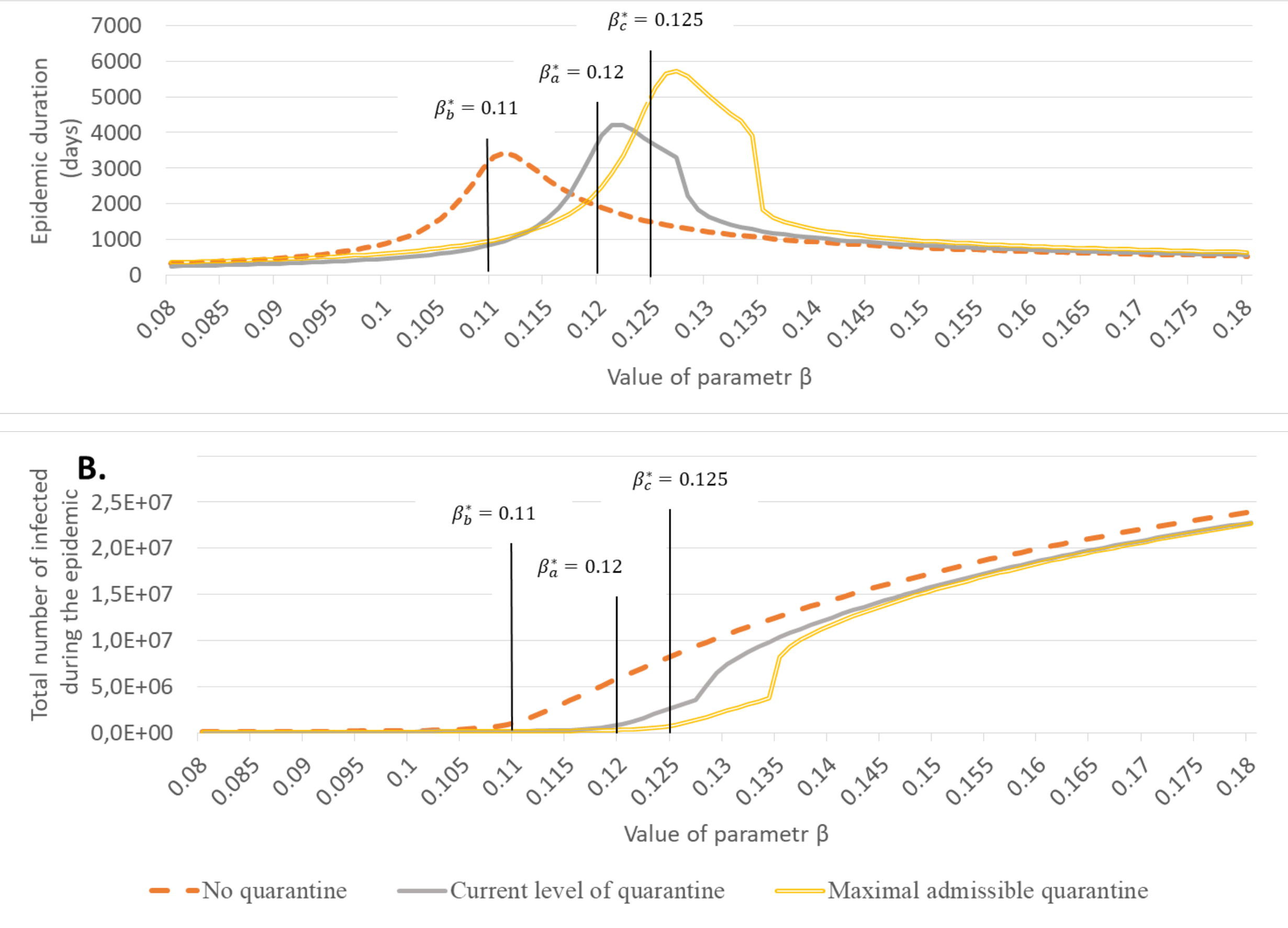}
\caption{Duration of epidemic and the final epidemic size as dependent on $\beta$, for $\kappa=0.2$.}	
\label{fig3.1}
\end{figure}

\begin{figure}[h!]
	\centering
	\includegraphics[width=14cm]{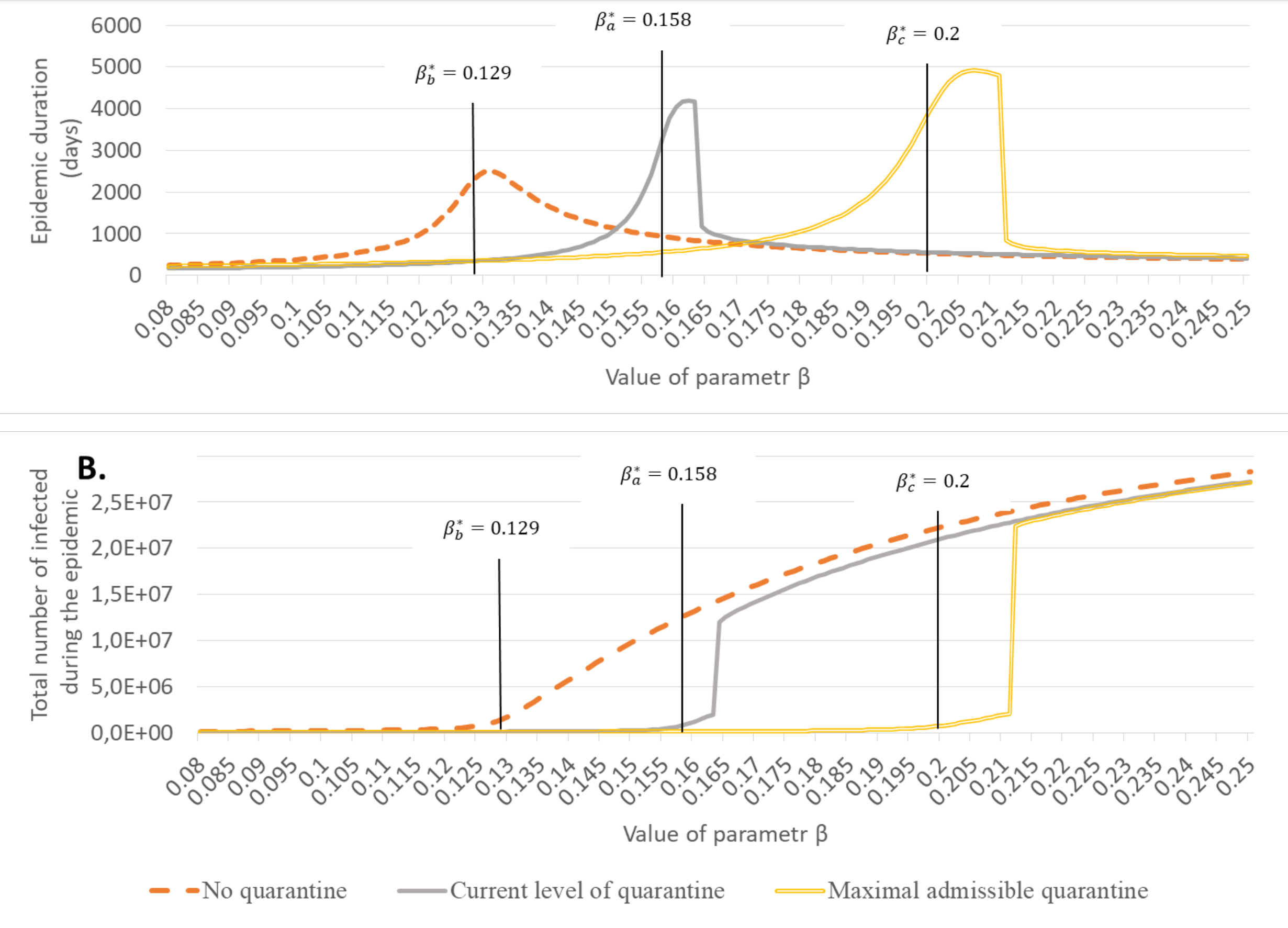}
\caption{Duration of epidemic and the final epidemic size as dependent on $\beta$, for $\kappa=0.5$.}
\label{fig3.2}
\end{figure}

\begin{figure}[h!]
	\centering
	\includegraphics[width=14cm]{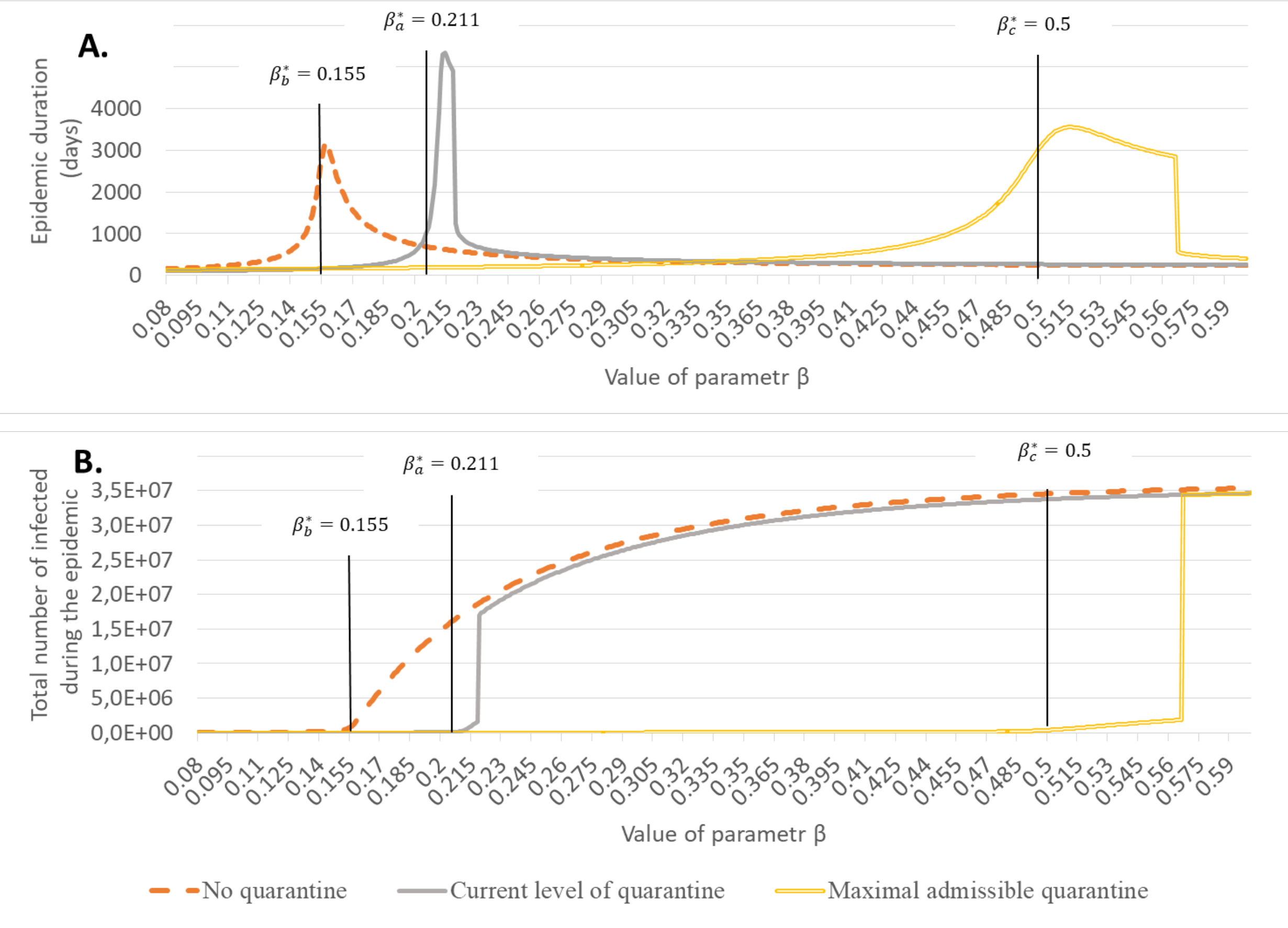}
\caption{Duration of epidemic and the final epidemic size as dependent on $\beta$, for $\kappa=0.8$.}	
\label{fig3.3}
\end{figure}

Our second goal is to simulate loosening of restrictions. In particular we want to verify numerically 
the critical thresholds $\beta^*$ listed in table \ref{tab1}. For this purpose we assume that at $t=60$
we change $\beta$. For each value of $\kappa$
we consider 3 scenarios:\\[2pt]
{\bf (a)} Current level quarantine: i.e. quarantine parameters $\theta=0.006, \; \alpha=75$
are maintained;\\[2pt]
{\bf (b)} No quarantine is applied starting from $t=60$; \\[5pt]
{\bf (c)} The maximal admissible quarantine is applied, meaning that  $\theta_{max}=\frac{\beta}{\alpha \gamma_d}$ (see (\ref{def:R0})).
In this case $\alpha=75$. As long as the limit $K_{max}$ is not reached there is no difference whether we increase $\alpha$ or $\theta$, the decisive parameter is 
$\alpha \theta$. Increasing $\alpha$ would lead to reaching $K=K_{max}$ earlier and hence worse outcomes.\\[5pt]  
Figures \ref{fig3.1}-\ref{fig3.3} show the final values of $R=R_d+R_u$
and time till the end of epidemic depending on the value of $\beta$ for $t \geq 60$. 
for above 3 scenarios and different values of $\kappa$. The theoretical values of $\beta^*$ are shown by black lines.

The results confirm that around $\beta^*$ a rapid increase
in the total number of infected occurs, coinciding with the peak total epidemic duration. Thus the numerical computations confirm that the critical $\beta^*$ calculated for the linear approximation in the section 2.2 are adequate, with a small bias towards lower values.

%
%


The case $\kappa=0.2$ shows that the influence of quarantine is not 
high, even for the maximal admissible case, when we are able to efficiently isolate all persons infected by every diagnosed.
%

 A striking feature in the behaviour of total number of infected
are jumps for certain critical value of $\beta$ observed for $\kappa=0.5$ and $\kappa=0.8$, 
both in case $\theta=0.006$ and $\theta=\theta_{max}$. The values of $R_d$ and $R_u$ before and after these qualitative changes are summarized in Table \ref{tab3.1}. 




\begin{table}[h!]
\centering
\begin{tabular}{|c|c|c|c|c|}  
\hline
 & $\beta$ & $R_d(t_{max})$  & $R_u(t_{max})$ & $t_{max}$ \\
\hline
& & & & \\[-8pt]
$\kappa=0.5, \; \theta=0.006$, & 0.163 & 1171 & 811 & 4170  \\[3pt] 
 & 0.164 & 6160 & 5875 & 1200 \\[4pt]
\hline
& & & & \\[-8pt]
$\kappa=0.5, \; \theta=\theta_{max}(\beta)$, & 0.211 & 1423  & 666  & 4800 \\[3pt]   & 0.212 & 11458  & 10971 & 840 \\[4pt]
 
\hline
& & & & \\[-8pt]
$\kappa=0.8, \; \theta=0.006$ & 0.218 & 1137 & 236 & 4740   \\[3pt] 
 & 0.219  & 13706 & 3365 & 1060  \\[4pt]

\hline
& & & & \\[-8pt]
$\kappa=0.8, \; \theta=\theta_{max}(\beta)$ & 0.566 & 1762 & 108 & 2850 \\[3pt]  
 & 0.567  & 27602  & 6729 & 570 \\[4pt]

\hline
\end{tabular} 
\caption{Critical values of $\beta$ obtained in simulations and corresponding final numbers of diagnosed/undiagnosed (in thousands) and total time of epidemic.}
\label{tab3.1}
\end{table}
A closer investigation for these values of $\beta$ shows that in all 4 cases the jump occurs for the first value of $\beta$ for which the limit number of quarantined, $K_{max} = 50 000$, is achieved. Notice that immediately after passing the threshold the values become very close to those without quarantine. Therefore the effect 
of quarantine is immediately and almost completely cancelled after passing the critical value of $\beta$. The transition is milder in the case $\kappa=0.2$ which can be explained by the fact that the transition takes place for lower values of $\beta$.

Results of our simulations confirm the theoretical prediction that the margin in relaxation of restrictions 
is very narrow if we want to avoid a blow up of the number infections. Strengthening of quarantine allows to remain in a stable regime while increasing $\beta$. 

\section{Discussion}
 We propose a simple SEIR-type model (SEIRQ), which includes the effects of testing and contact tracing. The model formulation allows to calculate an interpretable formula for the reproductive number $\mathcal{R}$ \eqref{def:R0}. As typical for this class of models, $\mathcal{R}$ depends on transmission parameters $\beta$. Increasing $\beta$ corresponding in e.g. to higher frequency of social contacts increases $\mathcal{R}$. Decreasing $\beta$, for example in consequence of widespread use of face masks, has the opposite effect. On the other hand $\gamma_d$ reflects the time to diagnosis and the formula indicates that more rapid diagnosis is associated with lower $\mathcal{R}$. In addition, our model offers a clear interpretation of the quarantine effect. The transmission rate due to diagnosed cases, $\beta_d$, is decreased by the factor $\theta\alpha\gamma_d$ indicating that both the number of quarantined per diagnosed individual ($\alpha$) and proper targeting of the quarantine (the infection rate among the quarantined $\theta$) equally contribute to this factor. Also the parameter related to testing: the delay in diagnosis, $\gamma_d^{-1}$, plays similar role. This quantifies the potential of a wide range of interventions to improve testing and contact tracing, as outlined in e.g. in ECDC recommendations \cite{ECDCCT}. In particular, as the number of people put in quarantine per each case and the infection rate among the quarantined impact $\mathcal{R}$ in similar fashion, our results support the recommendations to focus on the high risk contacts when the resources do not allow to follow all contacts.
 
 Our model takes into consideration only the effective contact tracing, i.e. the situation when the infected contacts are identified and put in quarantine before they become infectious. People who are identified later would be modelled as passing through one of the $I$ states to the $R$ states. This means that the number of quarantined in our model can be also increased by faster contact tracing. The timely identification of contacts may be a significant challenge in the quarantine approach given that the incubation time can be as short as 2 days in 25\% of cases \cite{Guan}. As mentioned by other Authors \cite{Ferretti}, the delays in manual contact tracing are usually at least 3 days and under such circumstances the contact tracing and quarantine alone may be insufficient to control the epidemic. This could be improved with digital contact tracing. Notably, mixed contact tracing strategies implemented in South Korea indeed helped to control the epidemic at the early stages \cite{Korea}. The use of "smart contact tracing" with mobile phone location data and administrative databases were also key to rapid identification and self-quarantine of contacts in Taiwan \cite{Chen} and implementation of such strategy helped Singapore to control the epidemic without major disruptions of social activities \cite{Ng}.
 
  We note that the quarantine effect relates only to transmission due to diagnosed cases. As expected, in order to control the epidemic the transmission due to undiagnosed cases has to be negligible. This can be controlled by general measures such as {\it lockdown}, which universally decrease the frequency of social contacts and are therefore likely to reduce $\beta_u$. In our model the part of $\mathcal{R}$ representing transmission due to undiagnosed cases is scaled by $(1-\kappa)$, the parameter relating to the efficiency of the testing system.  Again, the examples of Singapore as well as the Italian village of Vo’Euganeo show that the widespread testing complementing the efficient contact tracing was essential to suppress epidemic. Testing unrelated to epidemiological links decreases $(1-\kappa)$ factor, thus making the factors impacting transmission due to diagnosed cases, such as quarantine, more powerful to decrease $\mathcal{R}$. 
  
  Further, our model allows to study the effect of the situation, in which the contact tracing capacities are exceeded. In this situation the epidemic is likely to quickly develop to the levels observed without quarantine. It is therefore quite crucial to implement the aggressive contact tracing system, when the epidemic is still at low levels and it is possible to bring the epidemic to suppression phase.

We demonstrate the high impact of contact tracing and quarantine on the observed numbers of cases in Poland. This effect was coupled with substantial reduction in the transmission parameter $\beta$ resulting from social contact restrictions. Depending on the scenario, $\beta$ decreased by 76\% to 84\%, bringing $\mathcal{R}$ below 1.
The estimated effect of the quarantine in Poland would depend on which of the considered scenarios regarding testing efficiency was the most relevant to our situation. 
 In our model  the quarantine is estimated to be the most effective for the scenario in which most of the cases are diagnosed ($\kappa=0.8$). Testing strategies that comprise testing of all individuals with symptoms of respiratory illness could theoretically identify up to 82\% of infected, assuming they would all present to medical care. This could be coupled with random screening of high risk individuals, in e.g. health care workers, or - in case of high incidence - even random screening of entire community to achieve the $\kappa$ of the order of 0.8.  The Polish clinical recommendations specifically mention only testing all individuals with severe infections \cite{AOTMIT}. In addition testing is provided to health care workers. The severe course corresponds to approximately 18\% of all infections \cite{Guan}. Therefore, the $\kappa=0.8$ scenario is unlikely to be realistic in Poland. We believe that the plausible current $\kappa$ in our country lies between 0.2 and 0.5. For these scenarios the model shows that the control of the epidemic is largely achieved through suppression of $\beta$.  
In case of relaxation of social contact restrictions, the efforts should be focused on increasing the level of testing in order to decrease the proportion of undiagnosed cases as well as maintaining or increasing the effectiveness of quarantine. For smaller $\kappa$, even substantially increasing the effectiveness of quarantine does not allow to go back to the level of social contacts from before the epidemic ($\beta_1$).  

Finally, the contact tracing effort was manageable in Poland due to relatively small number of cases. Should the case load increase substantially longer delays in contact tracing would occur, which can substantially decrease the effects of quarantine \cite{Hellewell, Ferretti}.

\smallskip

{\bf Limitations and future directions of research.} 
 We do not consider the likely reduced transmission from undiagnosed cases who are more likely to be asymptomatic or paucisymtopmatic cases $(\beta_u<\beta_d)$. The reduction factor for infectiousness of asymptomatic is still under investigation. One study found a 60-fold lower viral loads in asymptomatic cases \cite{Liu}, but another estimated the transmissibility reduction by 50\% \cite{Li.R}. Moreover, we did not have sufficient data to include this additional parameter. We calibrated our model only to diagnosed cases, which were officially available. Calibration to mortality data is another approach successfully implemented in e.g. \cite{Flaxman} that potentially removes bias due to different testing policies. As there were relatively fewer fatalities in Poland and little data on clinical progression we decided on simplified model without explicit modelling of the outcomes.
 Furthermore, we did not consider the sub-optimal adherence to quarantine. It is likely that some individuals would not fully comply to strict quarantine rules. However, only anecdotal evidence for such phenomenon is available at this time. In our model it would decrease the effective $\alpha\theta$, which was chosen to fit to observed number of people put in quarantine.
 Finally, the analysis of $\mathcal{R}$ is suitable for small size of epidemic, when $S \approx N$. For other cases the results are still useful, but the approximation may be biased, as we have shown for $\beta^*$. 
  Due to little available data and policy changes we did not have sufficient data to determine which $\kappa$ scenario is the most appropriate.
 
 \smallskip
 
 In conclusion we present a simple model, which allows to understand the effects of testing, contact tracing and quarantining of the contacts. We apply the model to the data in Poland and we show that despite a substantial impact of contact tracing and quarantine, it is unlikely that the control of the epidemic could be achieved without any reduction of social contacts.

\smallskip

{\bf Acknowledgments.} This work was partially supported by the Polish National Science Centre’s grant No2018/30/M/ST1/00340 (HARMONIA).

\bigskip

\appendix
\section{Appendix}
\subsection{Optimization algorithm.} \label{sec:opt}
In order to fit the values $\beta_1,\beta_2,\beta_3$ we use a standard gradient descent algorithm. 
Namely, we define error function as 
\begin{equation} \label{def:fit}
\displaystyle E(\kappa)=\left[\frac{\sum_{t=1}^{54} [R_d(\kappa,t)-data(t)]^2}{\sum_{t=1}^{54} |data(t)|^2}\right]^{1/2},
\end{equation}
where $\{R_d(\kappa,t)\}_{t=1}^{54}$ is the vector 
of computed values of $R_d$ and $\{data(t)\}_{t=1}^{54}$ the vector of data (cumulative number of confirmed cases).

At each step we approximate the gradient of the error function 
with respect to $\beta_1,\beta_2,\beta_3$ by differential quotients and move in the direction opposite 
to the gradient. The algorithm reveals a good performance provided we start sufficiently close to the minimum,
which is not difficult to ensure in our case.

It remains to choose the initial data. A closer 
look on results of simulations shows that the choice of initial data mostly influence the fitting in the beginning
of period under consideration and hence the value of $\beta_1$, 
while for analysis of future scenarios $\beta_3$ is the most important. Taking all this into account we do not
struggle for sharp optimization of data fitting with respect to initial data and restrict to the following  
heuristic choice. It is natural to assume $I_u(0)=\frac{1-\kappa}{\kappa}I_d(0)$. 
Concerning the choice of $E(0)$ we assume it in a form $E(0)=m(I_d(0)+I_u(0))$. 
We set initial values $I_d(0) \in \{10,20,30\}$ and for each value we set $I_u(0)$ according to the above 
formula and three values of $E(0)$ corresponding to $m \in \{2,3,4\}$. For each of these $9$ combinations 
we run the optimization algorithm looking for the best fit of $\beta_i, i=1 \ldots 3$. We have repeated 
this approach for $\kappa \in \{0.2,0.5,0.8\}$. It turns in that for all values of $\kappa$
the best fit was obtained for $I_d(0)=20$ and $m=2$. More careful analysis around $I_d(0)=20$ did not 
improve the quality of fitting, therefore:
$$
I_d(0)=20, \quad I_u(0)=\frac{1-\kappa}{\kappa}I_d(0), \quad E(0)=2(I_d(0)+I_u(0))
$$    
is our final choice. We obtain the following fitting error defined by \eqref{def:fit}: 
\begin{table}[h!]
\centering
\begin{tabular}{|c|c|c|c|}  
\hline
& & &\\[-8pt]
$\kappa$ & 0.2 & 0.5 & 0.8\\[4pt]
\hline
& & &\\[-8pt]
$E(\kappa)$ & 0.0077 & 0.0079 & 0.0084 \\[4pt]
\hline
\end{tabular} \caption{Errors of data fit.}
\end{table}

\subsection{Choice of fixed parameters} \label{sec:par}

\smallskip
1. The parameter $\sigma$ describes the rate of transition from non-infectious incubation state $E$ into the infectious states $I_d$ or $I_u$. The median incubation time from exposure till the onset of symptoms was estimated at 4 to 5 days \cite{Guan, Li.Q, Lauer}. However, there exists evidence that typically infectivity preceeds symptoms, by 1 to 3 days \cite{Wei, Tong, Huang}. 

A modelling study identified the rate of transition between the non-infectious and infectious states at $\frac{1}{3.69}$ \cite{Li.R}, which corresponds to an average time lag of 3.69 days untill the case becomes infectious.

\smallskip

2. The parameter $\gamma_u$ represents the period of infectivity during the natural course of disease. We discuss the period of infectivity, especially as applied to mild cases. The median duration of viral shedding was estimated among 113 Chinese hospitalized patients. Overall it was 17 days, but it was shorter among cases with milder clinical course \cite{Xu}. A study among 23 patients in Hong Kong confirmed viral shedding longer than 20 days among a third of patients, although the peak level of shedding was noted during the first week of infection \cite{To}. In the mission report from China WHO reports viral shedding in mild and moderate cases to last 7 - 12 days from symptom onset. Among younger and asymptomatic or mild cases the shedding may be shorter: in a study among 24 initially asymptomatic youngsters the median duration was 9.5 days \cite{Hu}.

\smallskip

3. The value of $\kappa$ generally depends on the testing policy. However, recommended testing policies often rely on the presence of respiratory symptoms. This is also the case in Poland. 
It was observed that some infected people never develop symptoms, although the precised rate of such truly asymptomatic infections is still under investigation. Some studies may be biased by a too short follow-up time. A small study among residents of a long-term care skilled nursing facility found that even though more than half of individuals with confirmed infection were asymptomatic at the time of test, majority of them subsequently developed symptoms. The proportion of people who remain asymptomatic may be higher among younger individuals [Hu]. A study among Japanese nationals repatriated from Wuhan suggests the proportion of asymptomatic infections is about 30\% \cite{Nishiura}. An analysis among the passengers of Diamond Princess ship, where a COVID-19 outbreak occurred, taking into account this delayed onset of symptoms estimated the proportion of asymptomatic infections to be about 18\%, even though almost 50\% were asymptomatic on initial test. In addition, large scale screening implemented in Italian village Vo’Euganeo indicated that 50\% to 75\% of infected individuals did not report symptoms \cite{Day}. Similarly, in population screening in Iceland 50\% were asymptomatic at the time of screening \cite{John}. It may be stipulated that some of the people diagnosed through screening developed symptoms latter, consistently with the findings from the Diamond Princess study.

On the other hand a sizable proportion of infected people, especially at younger ages, experience only mild symptoms, for which they may not seek medical attention.  
In the study of Li \cite{Li.R}, the proportion of undocumented cases was estimated as 86\%.

\smallskip

4. The parameter $\gamma_d$ was estimated basing on a sample of case-based data available in routine surveillance, by fitting gamma distribution to the time from onset to diagnosis, for cases who were not in quarantine before diagnosis. 
Time from onset to diagnosis was estimated based on surveillance data available in the Epidemiological Reports Registration System for COVID-19, as of 28.04.2020.The system collects epidemiological data on cases diagnosed in Poland and is operated by local public health departments. All cases eventually are entered into the database. However, substantial reporting delays are noted.
There were altogether 4976 cases registered in the system, including 1995 (40.1\%), who did not have symptoms at the time of diagnosis. Plausible onset date and plausible diagnosis date were available for 2884 cases ( 96.7\% of 2981 cases that were not asymptomatic)

\begin{figure}[h!]
	\centering
	\includegraphics[height=10cm]{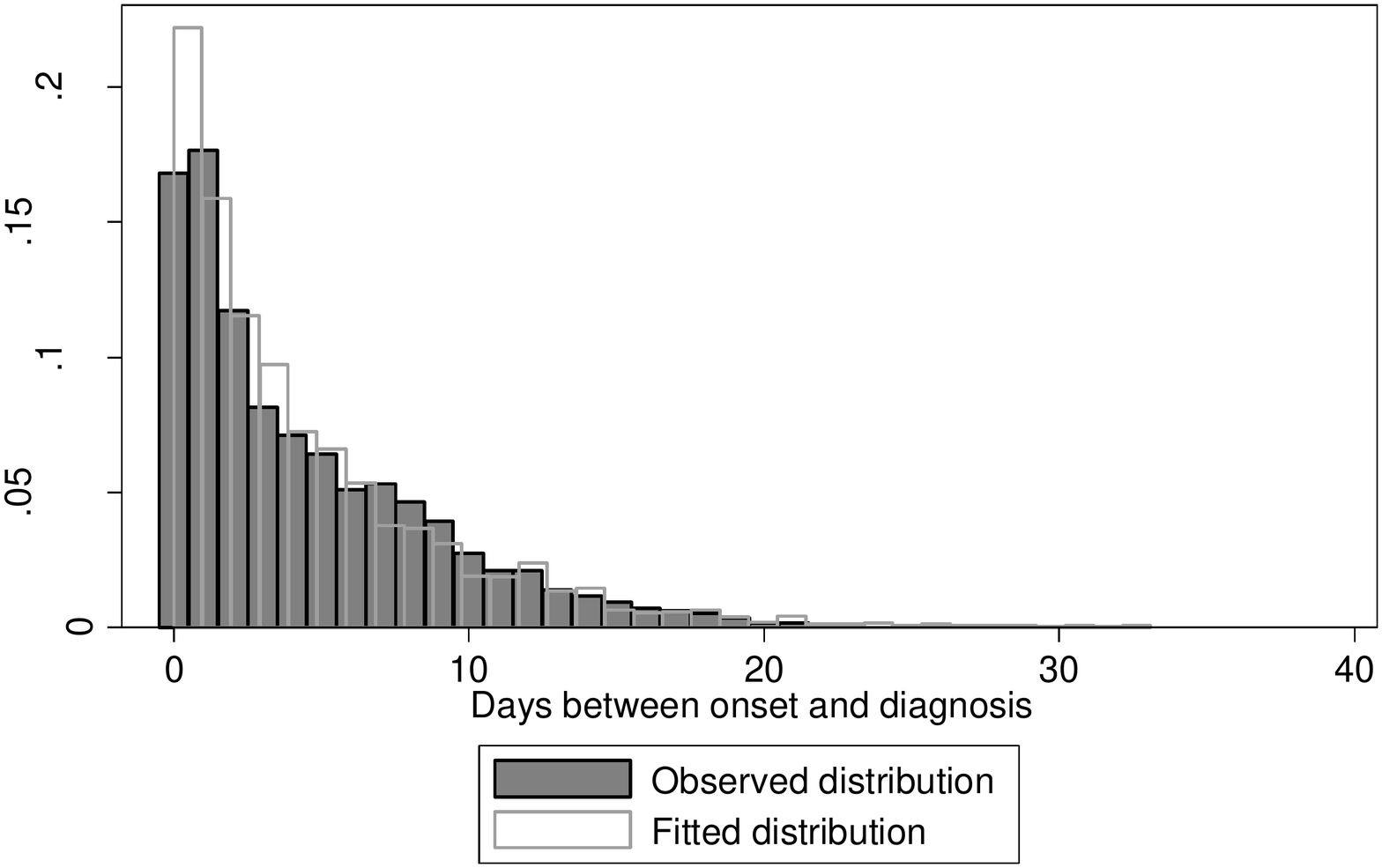}
\caption{Observed and fitter distribution of time between onset and diagnosis among symptomatic cases, who where diagnosed outside of quarantine. The fitted gamma distribution has the following parameters: shape = 0.91, scale = 5.06}	
\label{figA.1}
\end{figure}

Gamma distribution was fitted by maximum likelihood to cases who were not diagnosed in quarantine. The observed and fitted distributions are shown below (figure \ref{figA.1}). 

We next fitted gamma-regression model with week of diagnosis as an explanatory variable. We found no significant trend in time. 
We therefore adopted the average time from onset to diagnosis to be 4.6 days, and taking into account the probability of asymptomatic spread we assumed the parameter $\gamma_d$ to be 1/5.5.

\smallskip

\begin{figure}[h!]
	\centering
	\includegraphics[height=10cm]{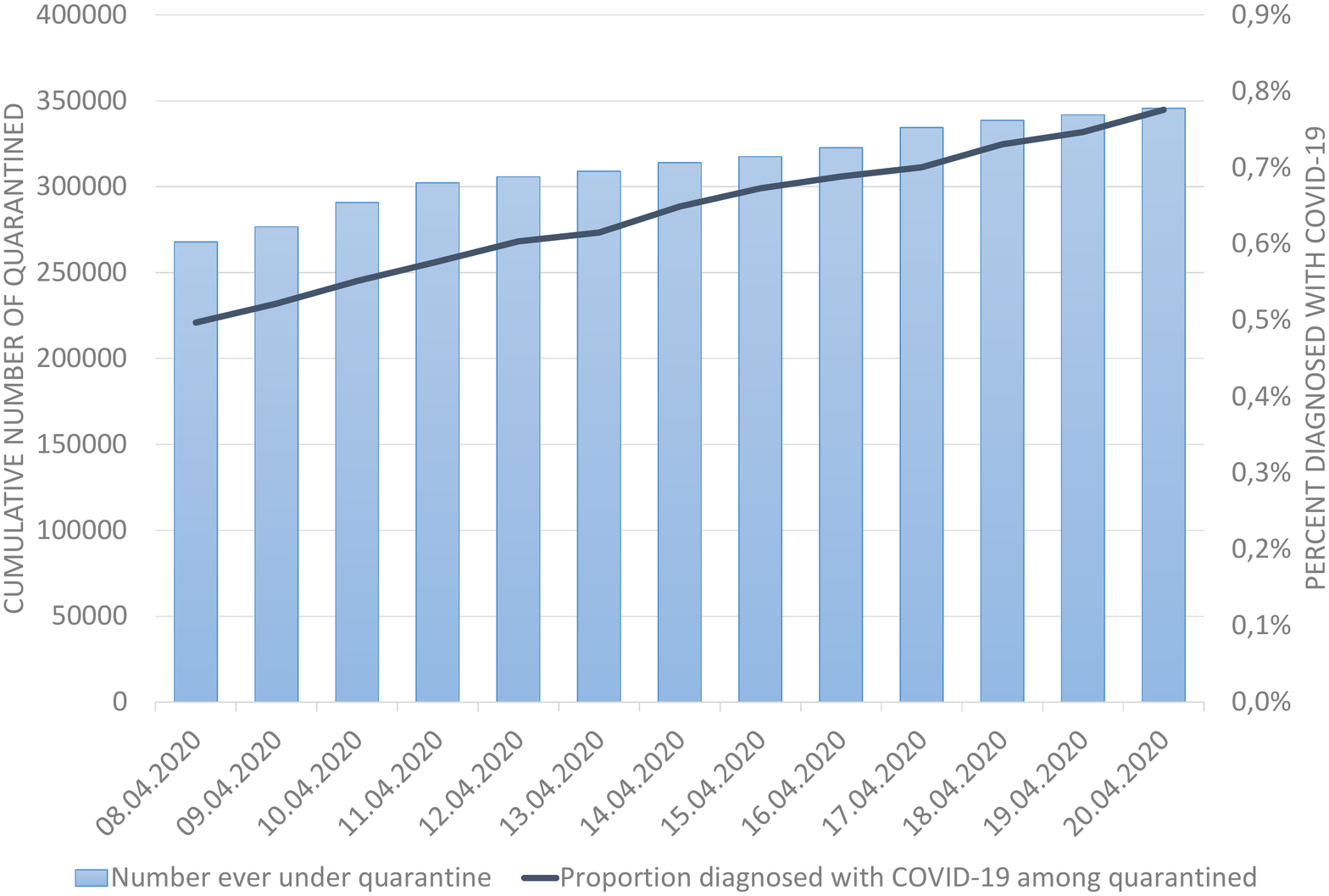}
\caption{Data on the population undergoing quarantine and the percent diagnosed with COVID-19 in this population}	
\label{figA.2}
\end{figure}

5. Next we base $\theta$ on available data. We calculate prevalence of infection among the quarantined individuals, according to data published by the Chief Sanitary Inspectorate on the number of cases diagnosed among quarantined people and the total number of quarantined. 
We used a series of data 8.04 – 20.04 to estimate a likely value of $\theta$. We chose this time period due to data availability. 
Data are shown on the figure below. During this time period there was an increasing trend in the proportion of diagnosed from 0.5\% to 0.8\% \ref{figA.2}. We presume that this parameter could change with changing procedures of contact tracing and testing. However, since no detailed data were available, for the modelling purposes we chose a simplifying assumption that $\theta$ is stable (i.e. we always take a similar group of contacts under quarantine) selecting an average value of 0.6\%.

 This proportion could be also viewed as attack rate among the contacts of cases. The proportion in Poland is in line with what was observed in Korea, where an estimated attack rate was $0.55\%$ overall and $7.56\%$ among household contacts \cite{Korea},
although household attack rate was higher ($>19\%$) in other studies \cite{Jing}. 

\smallskip

6. Furthermore we fix the parameter $\alpha$.  
Here we make another simplification assuming this parameter to be constant. The main difficulty is a lack of precise data concerning the number of newly quarantined people per day, distinguishing between reasons of quarantine (travel related or contact tracing relate). At the beginning of epidemic in Poland the average amount of quarantined following one diagnosed case was definitely higher. Moreover, people coming back from abroad were subject to obligatory quarantine starting from March 16 and constituted a considerable part of quarantined in the second half of March and beginning of April. In particular, around 54 000 Polish citizens staying abroad came back within a special program of charter flights operated by Polish Airlines which ended on April 5. We can assume that after this date the ratio of people coming from abroad among all people subject to quarantine was negligible. As our model does not take migration into account, we have to take into account only quarantine from contact. For above reasons, for fitting $\alpha$ we restrict our analysis only to a period of 2 weeks of April. Assuming already $\theta=0.006$ we then choose $\alpha$ minimizing the square error between the number of quarantined from the data and computed K(t). This way we obtain $\alpha=75$.

Taking these into consideration we set the values of parameters collected in table \ref{tab0}.

\subsection{Confidence intervals}
{\bf Bootstrap.}
To estimate confidence intervals we use a method of parametric bootstrap. We generate $M=200$ sequences of  perturbed data assuming that for each time $t \in \{1,54\}$ the increment of $R$ (i.e. daily number of new diagnoses) is a random number from Poisson distribution with mean value equal 
to increment of observed data. 

Model parameters are also perturbed, see below. For each series of perturbed data we estimate the values of $\beta_i$ and take 
estimated confidence intervals as appropriate quantiles of obtained sets. 

In order to estimate confidence intervals for $R_d(t)$, shown on panel A of figure \ref{fig2.0}, we proceed as follows. For each sequence of perturbed data we compute fitted $R_d(t)$. 
This way we obtain a set of curves $$\{R_d^{(k)}(t)\}_{k=1, \ldots, 200}^{t = 1, \ldots, 54}.
$$
Then for each time instant $t \in \{1,54\}$ we set lower and upper bounds of the confidence interval of $R_d(t)$ as appropriate quantiles of the set $\{R_d^{(k)}(t)\}_{k=1}^{200}$. Analogously we compute the confidence intervals for $R_u(t)$ shown on panel B of figure \ref{fig2.0}.  

\smallskip

{\bf Distribution of parameters.} 
Following other Authors \cite{Kucharski} as well as experimental data, for the uncertainty analysis we used the following distributions of the parameters.
\smallskip

1. $1/\gamma_d \sim Gamma(a_1,b_1)$, where the shape parameter,  $a_1 = 1.05$ and scale parameter $b_1=5.23$

2. $1/\gamma_u \sim Gamma(a_2,b_2)$, where the shape parameter,  $a_2 = 2$ and scale parameter $b_2=5$

3. $1/\sigma \sim Gamma(a_3,b_3)$, where the shape parameter,  $a_2 = 2$ and scale parameter $b_2=1.75$ 

4. $\alpha \sim Poisson(\alpha_0)$, where we assume constant $\alpha_0=75$;

5. $\theta$ - is not sampled for the uncertainty analysis. As the results depend on the quantity $\alpha \theta$, we rely on the distribution of $\alpha$.

We take $N=200$ (approximate average of daily number of diagnosed cases from the data). We approximate the mean value of $N$ samples from Gamma distribution using Central Limit Theorem. Namely, we generate
\begin{equation*}
\frac{1}{\gamma_d} \sim {\mathcal N}(a_1b_1,\frac{a_1b_1^2}{N}), \quad
\frac{1}{\gamma_u} \sim {\mathcal N}(a_2b_2,\frac{a_2b_2^2}{N}), 
\quad
\frac{1}{\sigma} \sim {\mathcal N}(a_3b_3,\frac{a_3b_3^2}{N}).
\end{equation*}

\subsection{Stability analysis - computation of $\mathcal{R}$} \label{sec:R}

Based on the classical approach to epidemiological models we address the basic question concerning the propagation of the disease. Namely, how many persons are infected by one infectious individual, a quantity which is usually called reproductive number, $\mathcal{R}$. In order to compute this quantity we use the approach from \cite{Dik}. 
We look at the system assuming $S \sim N$ and $E,I_d,I_u$ are close to zero, then we consider the 
following linearization
\begin{equation} \label{lin}
 \begin{array}{l}
  \dot E(t)= \beta_d I_d(t)+ \beta_u I_u(t) - \sigma E(t)  - \theta \alpha \gamma_d  I_d(t), \\ [10pt]
  \dot I_d(t)=\kappa \sigma E(t) -   \gamma_d I_d(t), \\[10pt]
  \dot I_u(t)=(1-\kappa) \sigma E(t) -\gamma_u I_u(t).
 \end{array}
\end{equation}
Note that in the above subsystem there are no delay effects. We write \eqref{lin} as 
\begin{equation} \label{2.2.1}
 \dot x = (T - \Sigma) x
\end{equation}
where $x=(E,I_d,I_u)^T$, $T$ is the transmission matrix and 
$\Sigma$ -- the transition matrix. $T$ has nonnegative entries and $\Sigma$ is lower triangular 
with strictly positive eigenvalues:
\begin{equation} \label{2.2.1b}
 T=\left(
 \begin{array}{ccc}
  0& \beta_d - \theta \alpha \gamma_d & \beta_u \\
  0 & 0 & 0 \\
  0 & 0 & 0
 \end{array}
\right), \qquad 
\Sigma = \left(
\begin{array}{ccc}
 \sigma & 0 & 0 \\
 -\kappa \sigma & \gamma_d & 0 \\
 -(1-\kappa) \sigma & 0 & \gamma_u.
\end{array}
\right)
\end{equation}
The system \eqref{2.2.1} can be rewritten as
\begin{equation} \label{2.2.2}
 \dot x=-(Id - T \Sigma^{-1})\Sigma x.
\end{equation}
Then one deduces (see \cite{Dik}) that if we define $\mathcal{R}=\max \{ \textrm{eigenvalues of \ } T\Sigma^{-1} \}$
then
\begin{equation*}
{ \mbox{ the system is stable for $\mathcal{R}<1$  and it is unstable for $\mathcal{R}>1$.}} 
\end{equation*}
Stability of system (\ref{2.2.2}) means that the whole vector $(E,I_d,I_u)$ is going to zero, it follows that the main system (\ref{seir}) also tends to the zero solution for $(E,I_d,I_u)$.
Instability implies that for "almost all" small data, the vector 
$(E,I_d,I_u)$ is growing in time (exponentially fast), causing 
the nonlinear system also evolves  far away from the trivial state, i.e. $E,I_d,I_u$ rapidly grow. 

By \eqref{2.2.1b} we have
\begin{equation*}
 T\Sigma^{-1}=\left(
 \begin{array}{ccc}
  \frac{\kappa(\beta_d - \theta\alpha \gamma_d)}{\gamma_d} + \frac{(1-\kappa)\beta_u}{\gamma_u} & 
  \frac{\beta_d - \theta \alpha \gamma_d}{\gamma_d} & \frac{\beta_u}{\gamma_u}
  \\
  0&0&0\\
  0&0&0
 \end{array}
\right).
\end{equation*}
Hence the stability of our system is determined by the following factor:
\begin{equation} \label{def:R}
 \mathcal{R}=\frac{\kappa \beta_d}{\gamma_d} + \frac{(1-\kappa)\beta_u}{\gamma_u} - \kappa  \theta \alpha.
\end{equation}

To make a final comment, let us note that in case of spread of pandemia, as $R_d,R_u$ grow, the above 
analysis become less reliable. 
Recall that $\beta$ is normalized by $N$, so as $S/N$ is not close to one and the analysis of stability becomes more complex. This behavior is illustrated by figures \ref{fig3.1}, \ref{fig3.2} and \ref{fig3.3}, where we observe deviation from predictions based on \eqref{def:R0},

\end{document}